# Piecemeal Journey To 'HALCYON' World Of Pervasive Computing : From past progress to future challenges

Rolly Seth
Amity University, Lucknow, India,
rolly.seth@gmail.com,

Rishi Kapoor
Indian Institute of Information technology,
Allahabad, India, rkapoor.rishi@gmail.com

Hameed Al-Qaheri
College of Business Administration,
Kuwait University, alqaheri@cba.edu.kw

Sugata Sanyal
Tata Institute of Fundamental Research,
India, sanyal@tifr.res.in

**Keywords**: halcyon model, pervasive computing, nonchalant, connected, hidden, calm technology, privacy.

*Although 'Halcyon' means serene environment which pervasive computing aims at, we have tried to present a different interpretation of this word. Through our approach, we look at it in context of achieving future 'calm technology'. The paper gives a general overview of the state of pervasive computing today, proposes the 'HALCYON Model' and outlines the 'social' challenges faced by system designers.*

## 1   Introduction

All the technological advancements in the world are the outcome of chimera of a handful that was seen much before. Technology has progressed by leaps and bounds in the past few years. As in the dream of Mark Weiser, 'Ubiquitous computing' [1] has started laying its foundation in today's real world, the future with pervasive computing assumes a pastoral world where an individual will not be distracted by plethora of information surrounding him. This paper encompasses our representation of this 'HALCYON' world and how will its piecemeal journey progress.

By saying the future world to be 'Halcyon', we mean a peaceful technology world, where although at any given instant of time, plenty of information will surround an individual, yet it will sit calmly without interfering with one's work or trying to grab its attention. A person unaware of its presence will perform daily tasks without the need to carry several devices wherever the person goes. Information will always be at his side and will come to his aid as and when the situation demands, like a 'guardian angel'. This might sound confusing as surfeit information always distracts users, but this is the charm of pervasive computing "Everywhere, Anytime yet ulterior".

To realize this, it requires a shift in view from a sender's need to that of the receiver's. For example, our mobile phone rings whenever the sender wants to talk to us, irrespective of our preoccupancy. Although we have the option of picking up the call or disconnecting it, yet it asks for our attention at that instant only and thereby interrupting us. Thus, this perception is far from the realization of the halcyon world. We require technologies like e-mail that will be present at the receiver's end but only be seen when the user is at ease and wants to use it.

This paper gives a general overview of the current state of pervasive computing in section 2, proposes the 'HALCYON Model' in sections 4 and 5 and outlines the 'social' challenges faced by system designers in section 6.

## 2   Background

The clear indication for pervasive world to come into existence is through the use of "asynchronous communication". The reason why today's world cannot be categorized into that of 'calm technology' [2] is that a sender usually requires immediate attention of the receiver. Both the sender and the receiver are required to be present and free at the same instant, which is not always the case. In 'Halcyon' world, information is only provided to the user when he wants it.

A vital consideration while developing such a world requires each device to perform varied functions corresponding to different environments. In other words, devices should be made 'environment specific' and not function specific. This is in contrast to today's world where a user's mind quickly maps to one particular device, being subjected to a particular requirement. For instance, to measure human temperature, one would look for a thermometer and so on. All this boils down to the use of 'augmented functionality', that is add-on functions to be performed apart from the basic ones.

```
H: Hidden
A: Adaptable
L: Language Independent
C: Connected and collaborating
Y: Year-long
O: Openness
N: Nifty and nonchalant
```

Figure 1: meaning of HALCYON

We conceptualize the meaning of 'HALCYON' as shown in Figure 1:

**Hidden**: The first and foremost requirement of a peaceful/halcyon world would be that devices and their working be concealed from the outer world. By this, we do not mean intrusiveness into one's life. It would be of non-intruding type, thus safeguarding the privacy and safety of an individual. This is a major issue against fast pace progress towards a 'calm technology' world as to how such surplus amount of information will be kept hidden and administered so that it is not misused? For this to happen, the invisible front-end is to be supported by complicated back-end servers.

**Adaptable**: It should be easily adaptable to the environment or be context-aware. It should have the flexibility to re-configure itself. It governs the display properties of the information. For example, if someone calls the user and he is busy driving a car, then the environment will sense that and divert the call to another device, an answering machine or so. A premature example of it would be the popularity of call diversion to voice messages in today's era, yet it requires more refinement to adapt to the bigger vision. It requires 'diversification', that is to choose a device from a set of many others that best fits the requirement instead of any predefined diversion route.

**Language Independent**: By language independent, we do not mean non-standardization of platforms. On the other hand, we mean that halcyon world should not be restricted to the use of only one mode of communication, which is through use of words. It should involve other modalities too. The environment will decide which one of them should be used according to the situation. For example, the use of visual representation and colour changes are the most common options used today as in datafountain [3], Auraorb [4], flashbag [5] etc.

**Connected and collaborating**: This is the major role after being 'hidden' that devices have to perform. Physical distance would not be a barrier to connectivity. No doubt mobile phones offer this paradigm but for one to stay connected even at remote places, they need to answer a call and talk or reply via SMS etc. But, in contrast, in the HALCYON world, one does not need to disturb the other person in order to stay connected with him. Consider for instance, that the wall of our room changes colour according to our work and design patterns which depict our mood. A blue wall with flowers on it shows that we are in a cheerful mood. Some ambient interfaces like hello-wall, forecast umbrella and visual calendar have shown a promising step towards this development. Through connectivity, one could even be able to control the operations of their home appliances from remote places.

Connectivity would also require mutual collaboration between the devices. A message will be automatically transferred from one device to another without the use of human intervention. Consider one such example of an alarm clock, which apart from waking up an individual will also request the kitchen to make tea or breakfast for a user and kitchen in turn will look at user's past history, interest, health report and other details to choose the best breakfast to be made. It will also check with the health report if the user is ill or not. If yes, then it will instruct back to the alarm clock so as not to ring and wake the user as he is not well and waits till he gets up. If his reports are fine then kitchen will prepare breakfast by collaborating with kitchen appliances by the time the individual gets ready to go to the office.

**Year-long**: Back–end servers will need to be run continuously, gathering 24X7 data from different inputs and using this data for future use. For continuous usage various constraints come in, like enormous power requirements for day/night running. One solution available for continuous use is the usage of RFID (radio frequency identification) [6] tag/reader used in wearable computing. Polaris [7] being one such example.

**Openness:** It means open access to data or material resources. Any one should be able to extract information quickly from the 'open ocean' of data and use it for own good. The biggest reason for Linux platform's fast popularity was it being available as open source to one and all, where anyone can use and change to meet one's requirements. For this to be possible a large 'repository' needs to be maintained. One such approach is PIE [8], (Personal Information Everywhere). It will be a boon in various ways. Consider the case of healthcare, where information stored in the past would be used to cure an individual at present and provide medications whenever required. Many wearable and ambient displays are made for healthcare purposes which show blood pressure, sugar level and according to past history, warn the user in case of any danger, but that is just the beginning. Openness of data will also help in removal of problems like traffic congestion and will automatically command the driver to take another route in case of traffic congestion on a certain route. The hindrance to this paradigm is obviously how to prevent the sensitive information from falling in wrong hands? Balancing between full and legitimate usage is an issue yet to be tackled with. Also storage of such surfeit information still remains a question and 'Memex' [9] being one such proposed concept. There are further serious limitations on memory usage.

**Nifty and nonchalant**: Although it would be very complex making such an environment, it must be easy to use and understand. It should naturally blend with the surroundings so that an individual does not feel disturbed by its presence. It must also be nifty to understand the changes in the surroundings and "dynamically" adapt to them. Some companies like Ericsson and Electrolux are developing an intelligent refrigerator [10] to detect shortage of any food item and automatically order it from the supplier. To some extent, intelligence has already been inbuilt into today's washing machines that have the power to detect water level needed, soap requirements etc and automatically rinse the clothes and then switch off the power supply. One needs to develop such adroit machines that would sense a danger and then automatically work to revert it. Although natural disaster sensing machines have been developed, their functions do not extend beyond displaying warning signals. For a halcyon world to be developed, 'integration and synchronization' of many tasks should take place. Like a fire alarm on sensing fire in the house will automatically dial and call the fire control department and before the rescue team arrives, will try to extinguish the fire by itself by turning on the sprinkler valves.

## 3. Related Work

Pervasive Computing is the envisioned fifth generation of computers which has evolved through a piecemeal journey of seventy years when the first generation of computers came into existence in 1940s. Since then each leap to the next level brought many drastic changes for computers. And so this is expected from this new wave of computing. Among profuse metamorphosis in various areas in each era, we would like to give a special mention to:

1->Size,
2->Memory (Data storage) Capacity,
3-> language for interaction and
4-> Input / Output Display

These became the benchmarks for assessing the technological advancements of each era.

The first yardstick of evaluating development had been reduction in size. The first generation computers used to occupy the whole room which UNIVAC and ENIAC belong to. Then up to the fourth generation, a number of alterations were done through the use of Integrated circuits. A large number of components of the computer could be placed on a single Silicon chip, the size of a finger tip. Moore's law still holding well, what could be prophesied is that the size would become so small that it will touch atomic encode information. Also the much awaited use of 'Nanotechnology' provides flexibility of doing things that could not be thought of earlier. The new promising 'Nokia Morph' [11] makes use of this nanotechnology to provide flexible, transparent, context aware mobile devices.

The second yardstick is information storage which has always been a concern with ever increasing requirements. Actual size of devices is decreasing, yet, there needs to be steady increase in the memory size. The initial concept in this field was the 'Memex', put forward by Vannevar Bush [12], about a large place to store and modify data as per user demand, which laid the foundation of continuous progress in this area. Initially, magnetic drums were used to quench the demands. Then came technologies like floppy disk, magnetic tape etc for secondary data storage. Presently, semiconductor memories are much more prevalent. However, this change is also an interim solution. In order to hold much more data again, nanotechnology is going to play an important role. However, several other software concepts are coming into picture, not only for storing more data but to store what we call 'Intelligent' data. 'Conceptnet3' [13] is one of such ideas which adds common sense knowledge to the data being stored. Thus, there has been a shift in functionality now, as to what else a memory could do apart from storing data, thereby trying to match it with human brain.

The third benchmark is the language required in order to interact with the computer and get its work done. The transition of language has been judged from the ease of use for the user in commanding the computer. Each generation has shown an increasing level of abstraction from earlier 0/1 binary machine code and mnemonics in assembly language to today's English like sentences used in High Level Language. Still the need for a 'natural language' is felt where one does not have to memorize the instruction set and rules to be followed while directing machine to perform a certain task. This is what has emerged as Artificial intelligence languages for fifth generation languages such as LISP and PROLOG [14]. LISP, due to its inbuilt structure takes less effort and less memory space to do a similar task as compared to ADA or any other language. However, it is still far apart to achieving a natural language tag which could be used for pervasive computing.

The last yardstick but of paramount nature in context to pervasive era is the use of Input /Output devices chosen for gathering data from environment and displaying information in the environment. The initial generations relied heavily on use of punched cards, printouts. Then came GUI (graphic user interface) and mouse for I/O. Mark Weiser described such devices namely Tabs, Pads and Boards in his Pervasive Computing vision. Some new era devices making use of it are Ambient Trolley [15], Aura Orbs, Infocanvas [16] etc. Nowadays, apart from taking the information from words, other modalities have also been explored. What is emerging as the unanimous choice for making pervasive I/O devices is the use of RFID (Radio Frequency Identification) where tags attached to different objects send radio frequency in all directions (each tag is associated with unique frequency) and accessed by readers who want it at that instant. It implements the pervasive concept that information will be floating around the user any time and he will access it whenever he feels the need of it. Another gadget being widely used nowadays is 'Phidgets' [17], [18] that can be attached to the USB port of a PC and acts as a building block for providing sensing and controlling operation from the PC. It is built with a powerful Application Programming Interface (API) which provides the user abstraction from the underlying working. It can sense various functions such as light, distance, humidity, motion, pressure, touch, voltage etc and can be interfaced from a wide list of platforms like Java, C, C++, MATLAB etc.

It is worth noticing here that with each wave of new technology more stress is given on visualization. The most recent archetype of it being Hello-wall [19], Data fountain and History table cloth [20]. These make use of daily life objects to display information. The most exciting point of these inventions is that the information is visible through them 24X7. Yet these technologies do not interfere with one's daily routine to grasp their attention. Data fountain will display the real time comparative currency rates of different countries by means of the height that water level rises to, in each of the fountains (assigned to each currency). Similarly, Info Canvas provides information with the help of paintings; such as, number of people in the painting will represent the traffic conditions on the road; Sky colour of the painting will show weather condition in the real world etc.

As each era progresses from one to another, it marks the beginning of some new modalities. Pervasive Computing aims at not restricting itself to only one label. Any mode will be chosen from one of the several others that are available, as and when required. Thus, with today's technology it tries to explore every possible mode of providing information - be it voice, colour, area, picture or sound.

## 4. System Overview

### 4.1 Pervasive Computing: Other side

Although everybody expects Pervasive computing to be a boon for everyone, it has a graver side of it too. As they say 'everything comes at a price and so it has'. In this paper, we also try to unfold this other side of the coin. The effects of pervasive computing can already be seen in the social life of us.

Primarily, artificial intelligence will result in automatically cutting down physical activity of an individual, who will now not get up and walk much, in order to perform the daily routine tasks. We assume that devices will be integrated to pass on a request from one to another by themselves. This will, in turn, have an effect on biological metabolism which requires certain daily dose of physical activities for its proper working. A computer animated movie 'Wall-E' (2008) also tries to portray this issue where people have become boneless as they do not even need to walk in order to get their work done (as some machine does it for you, right from brushing your teeth).

Huge amount of e-waste [21], [22] would be generated, and disposing it could become a problem.

Circumlocutory, the radiations from it will affect the human body resulting in various obscure diseases or defects.

Social interaction will subside as people will not feel the need to talk to one another to know their well being (as that information will be automatically sensed) or to get some work done. We human beings, as social animals will thereby have drastic changes in our behavior, who will then consider computers as better companions than Homo sapiens.

Dependency on computers would increase, as humans will feel helpless in performing any task without the availability of machines.

Indirect effect of decline in human-to-human interaction will increase aloofness and probably more chances of suicide rates as then physiological counseling of one (provided by talking to family, friends etc) will be less and surfeit of information will even guide different ways to commit suicide.

Risk rates can become enormously large as a small mistake in commanding a machine or interpretation by a device can lead to severe damage. This may result in a series of many others that are collaboratively working. Thus, a small aberration will take the scale of much bigger mishaps.

Last, but of paramount importance is the concern for privacy, as anyone and everyone's sensitive data will be floating all around. An individual would be easily tracked or others would be able to see which activity you are currently involved in. Similar to the problem of e-waste, there will be issue of spamming. Our email in-boxes are already filled with those. It is not difficult to imagine what will happen when these seamless machines will be monitoring each of our activities.

Also consider for example project Aura [23]. Carnegie Mellon University conceived of this project as an example as a grand ubiquitous computing project, aiming to have a large scale computing system, demonstrating this concept. This one covers wearable, handheld and other type computers. Pervasive computing system keeps track of user location, behaviour and habits. It constantly tunes itself on basis of above data. Though this information is necessary for successful system deployment, it is a serious threat to user's privacy. It is important to question our design principles of the "pervasive computing system" so as to strike a balance between the "invisible machine" and loss of user's privacy.

### 4.2 Pervasive Computing: Benefits

However, apart from all these shortcomings, following benefits will always be provided to one and all through pervasive computing: In Halcyon world we can imagine a world with easy access to information resulting in time saved to obtain that information. Consequently, travel costs required for getting the work done will decrease.

A transparent society [25] would be created where anyone will not be able to provide anyone with wrong information. Your family and friends will know about your well-being without disturbing you again and again by calling or messaging. A huge number of accidents could be dodged at the correct time thereby cutting down the enormous premature death rates due to mishaps. Traffic congestion problems could be avoided. Help will be at your doorstep as and when required to assist you in case of emergency.

One does not have to search for the appropriate information required for his use out of the plethora of information. Smart devices will automatically find it for you. Things will be linked together working collaboratively thereby eliminating the need to feed output of one machine to another for further processing. Their intelligent system will self regulate that.

## 5. Methodology Proposed

Since the idea of pervasive computing was first proposed, much has been written on it but still there is a need for a single unified approach for integrating all pervasive computing. As can be seen in figure 2, the whole model is conceptualized by using a 2-channel grid technology. The two channel system will aid in duplex communication where one channel will allow anyone to components. Through this paper we will try to propose a 'HALCYON MODEL' in order to pace up the process of continuously transmit information and other one being for simultaneous reception. These two channels will cover the entire globe using grid system.

An individual will then send or obtain information from any of these channels without disturbing anyone in the process.

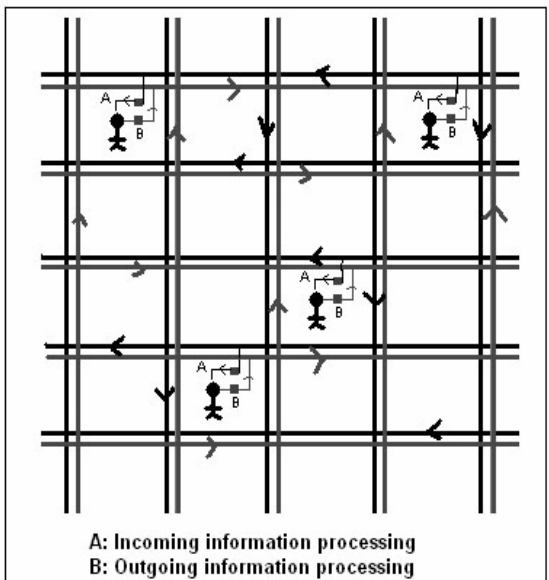

Figure 2: HALCYON Model of pervasive world

The prime focus of this model is on the two small boxes labelled A and B in figure 2. These two processing subsystems will include all the main considerations required for calm technology world to come into existence. Figure 3 shows an enlarged version of box A which will come in between an individual and the entire grid of continuous flowing information. Whenever this individual will consciously or unconsciously need any information to his aid, he will extract information from this free flowing ocean. Before it is made accessible to him, it will be needed to be processed inside subsystem B to check whether he is an authorized user. As can be seen in figure 3, seven levels are to be passed after raw information is read from the grid. Let us see each one separately.

**LEVEL I - SECURITY CHECK LEVEL:** This level is to address the most important issue hindering the fast pace development of pervasive world. Here an individual identity will be checked if he is the legitimate user to have access to the specific information or not? One of the many approaches could be through the use of RFID tags/ readers system where each individual would be provided with a RFID tag. One must note that with the free flowing information in the grid, a list of authentic users also run along with it to check that the validity of an individual with that list. It will therefore provide privacy in a way that even though the information would be flowing around everyone, only appropriate users will be able to use it. If the user stands justified obtaining information, the information is directly sent to Level VI.

**LEVEL II- MODALITY AND DEVICE DECISION ACCORDING TO URGENCY:** The second level considers the fact that in HALCYON world, the things have to be context aware or adaptable to the environment. In order to make this possible flexibility needs to be provided regarding the selection of device to be used for providing the information. In this level firstly urgency level and environment is checked to know how quickly the data is to be provided to the user so that it will aid him at right time. If user is busy now and it is found inappropriate at the moment to provide the information to him, it is sent to a queue to be passed on to someone else. Otherwise, different modalities are chosen according to the priority levels like vision, sound, smell, feel etc. If the information is not important for the moment then it is also sent to a queue and stored for future use, so as not to disturb the user at the moment

It is worth mentioning here that it purely depends on the kind of information as to which path to be followed. It does not solely depend on urgency level sent. For example, fire at home (urgent) could be tackled by someone else if the user is currently busy in a meeting. But urgent information like a chance of accident with a truck while you are driving is to be provided at the same instant. Thus, you can see that although both pieces of information were urgent, yet their processing was different depending on user environment. If the information to be provided to the user is important at the

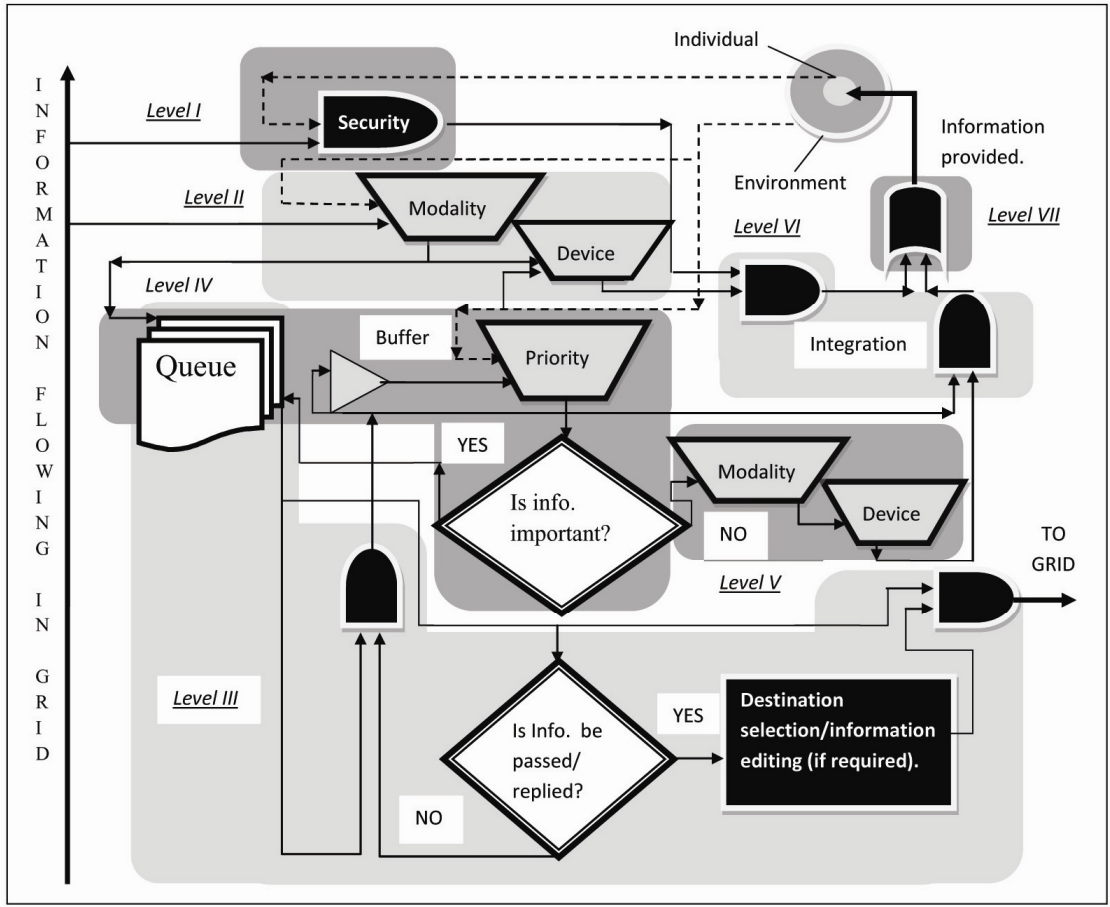

Figure3: System 'A' of HALCYON model

given moment only, then one of the modalities like sound, vision, touch etc is chosen, whatever is more appropriate at that instant. After the choice of modality is done, this and environment data are taken as selection lines in order to decide the appropriate device available with the user like cell phone, speaker etc in his environment, to pass on the information. This accounts for the fact that one has the option to provide information only through available devices with or around the user.

For example, consider you are driving a car towards north and a truck is coming from the right lane. There is high probability of a collision between the two vehicles. According to urgency level of decision, it is decided that the information of utmost importance is to be provided to the user at that instant. Considering the scenario, the appropriate modality will be voice. From the environment, it is found that speaker is the best device available with the user in order to draw his immediate attention. The user can then take immediate action to prevent the same.

**LEVEL III- DECISION TO SEND INFORMATION TO OTHER SYSTEMS/ REPLY BACK FOR FURTHER OPERATION:** At several occasion it becomes difficult when systems/devices demand your attention in order to send their output to the next one for further operations to be done. Consider a case when you are busy in a very important meeting in your office and your house catches fire. Although it is very urgent, you can not give immediate attention to it. Now, a possible solution offered in the pervasive world is that accessing your preoccupancy does not disturb you. On the other hand, it automatically gives authority to the fire extinguisher to switch on until the fire has been extinguished. At the same instant it must also call fire department if it senses that this fire is of massive scale which can not be put out by extinguisher alone. Thus, a decision is made at this level to know if the individual did not receive that information since it is not urgent or user is very busy in some other work. In this case, the first option is found correct and information progresses to Level IV else information is sent to the grid with proper authority list of users and urgency level. If the information is to be passed on, firstly the information is deciphered to find out the appropriate users where this information is to be sent, depending upon some predefined set of criteria. For example, in this fire case, fire department is found to be the appropriate receiver. Also, reply is given according to some predefined rules.

**LEVEL IV- UNIMPORTANT INFORMATION IN QUEUE RECHECKED FOR PRIORITY AFTER A CERTAIN TIME DELAY:** In case the information is not required at that instant, the user is not disturbed and thereby sent to a queue, which is unique for every individual. In case there is no hurry to send that information to other system/user then after a certain time delay, again the data stored is checked in "First In First Out" (FIFO) order to decide for the fact if it still is of any value to be provided at that instant or not? If not, then data is sent back to the queue again. Else it is sent to level V for further processing. One must note that the queue portion is common for both Level III and Level IV usage.

**LEVEL V- CHOICE OF MODALITY AND DEVICE OF DELAYED INFORMATION ACCORDING TO URGENCY:** If the data is important for the individual and this is to be known now, not after certain time, the same process as explained in Level II is repeated. Only difference being here is that now the option for sending data back to queue is not provided as now the information is of value only for that instant. Actually one must have seen the familiarity of these blocks with the digital world where such symbol represents a multiplexer. Thereby only one output is chosen among a plethora of many others depending upon one or more select lines.

**LEVEL VI- INTEGRATION OF INFORMATION WITH CONTEXT AWARE DEVICE SELECTED:** This level includes two parts. First being merging of the information with the device selected from those available at that instant only. Second also performs the same operation but it is kind of delayed version of the same coming from the queue.

**LEVEL VII- INFORMATION MADE AVAILABLE TO THE INDIVIDUAL:** This being the final level, here the information is coming from either of the two separately integrated portions and is made accessible to the individual at that instant only. Thus, one can see that this 7 levels subsystem A could to be installed between the free flowing information grid and an individual / device in order to transform this world into a calm one.

The proposed framework so far focuses only on one part of the whole picture, that is, how the information will be accessed by someone. The second part encompasses how it will be sent to the grid. Although, in part A, some information needs to be sent to the grid, it is for the sole purpose of linking two processes or operations. In other words, there is just an intermediary information link. On the other side, the subsystem B will depict how the information will be delivered if it directly originates from the user /device with no predecessor. However, one must note that operation wise the two portions are similar. Figure 4 shows the subsystem B of how the originated information is sent to the grid. It is quite simple in approach compared to that of A. Here the only fact to be kept in mind is that the information can not be simply sent to the grid of free flowing information as then various security and destination issues will pop up. Thus, in order to prevent that from happening along with the information, three other details are also sent along with it. One is the identity of real destinations, only where the information is to be made accessible, identity of the sender and the urgency level with which it is to be provided at the target. Figure 4 accounts for these facts only.

Maybe one should not overlook the fact that the focus has been shifted to the receiver rather than sender which demands more processing at receiver end. One could reason out that generally, everyone who sends the information will send it on high priority, to be processed at that time only. But, the model, being receiver centric, data is sent without giving preference to its urgency. Thus, if the user is busy, the information is sent to a queue where it is either sent to some other user, to be processed immediately or checked after some time if individual is free and it is necessary to provide the information. One application of it could be to stop numerous advertisements that demand your immediate attention.

Overall, it must be kept in mind that all these levels are only broad classifications not restricting to only one operation each level. Also, any level can be skipped, depending upon the requirement.

It should be kept in mind that these two subsystems namely A and B will be ulterior and continuously running in the backend without one's knowledge.

In order to give a better picture of the model proposed let us consider an example. Here we have tried explaining part A by taking the fire instance. Imagine one house has caught fire. Now, as shown in the figure it immediately sends information to the person X (to whom this home belongs) by embedding it in the free flowing channel. The information sent contains four parts- Message: 'Fire at home', sender, receiver identity, and urgency level. This information sent reaches the person X but before being accessed by the user, it passes through subsystem A. The processes happening at each level have been explained alongside. It is to be again mentioned that all levels have not been covered in it as there was no need for those in this scenario. At level I, firstly security check is done if the given user is person X or not? Since if he is found to be the legitimate user, level II is then followed where it is initially checked if he is free or not? If free then appropriate modality and device is chosen for providing information. In this case as seen from the figure 5, the user is busy in a meeting and thus not free. Thus, this information is sent to the queue to self manage the things, which transfers it to level III. At level III it is checked if information is to be passed on to other systems by providing authority to them. Since, fire at home is of utmost priority it is passed on to fire station (after deciphering the message that there is fire at home, so fire department is to be called according to predefined actions). Apart from it, it also replies back to home to use fire extinguisher in the meanwhile, or call neighbours. In case it is difficult to decipher the message by the system, it can be passed to some other individual, for example secretary or some other relative.

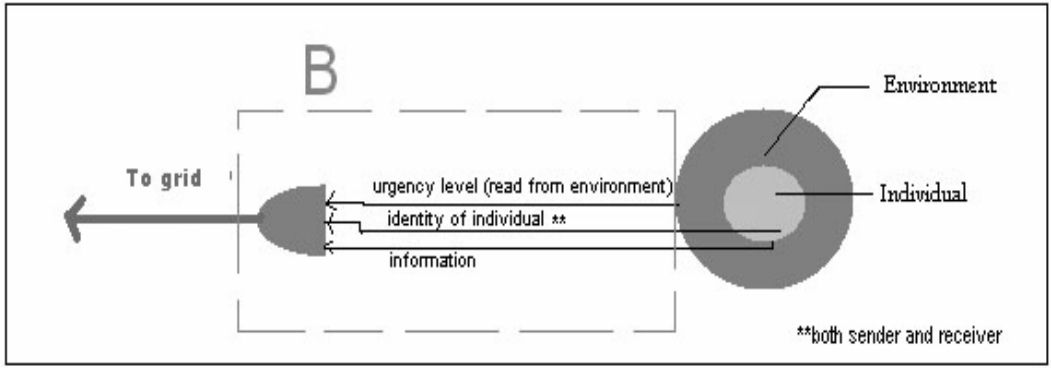

Figure 4: Subsystem 'B' of HALCYON model

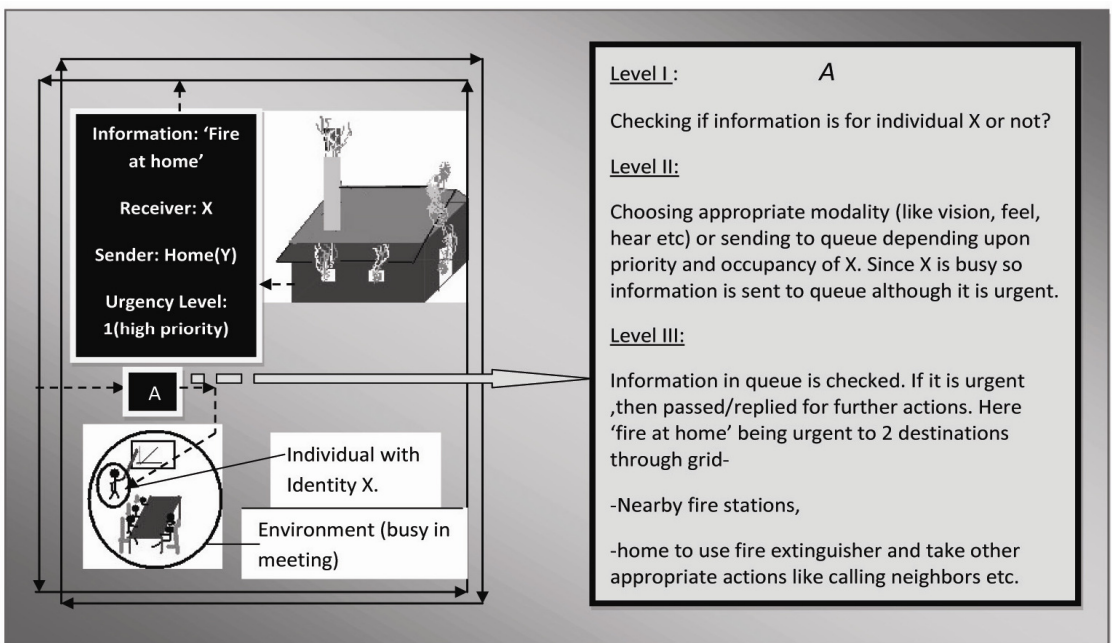

Figure 5: A sample case of halcyon model.

# 6 Issues and Future Challenges

The major issue that still needs to be figured out in adopting this model is to discover a technique through which such huge amount of information could be stored, which is exclusive for each individual (as shown in figure 3, where each individual is provided with his unshared Queue).

As 'HALCYON' will be for the whole world, the whole world should unanimously agree upon a common standard such as Sync ML, XML or similar proposed standards [27]. Other challenge is of finding a universal standard for the communication and the flow of this information in the grid. Although many proposed solutions are available, everybody has not agreed upon a single one method. In case two systems co-exist like Bluetooth and 802.11 [26], they will result in interference, thereby affecting proper working of each other.

Another concern is of finding a way for spectrum allocation so as to provide easy access and identity to every device/individual as spectrum shortage will be a process considering the mass scale of such operation. An idea proposed is to keep monthly subscription for everyone who only can access the portion of spectrum provided he has paid their monthly charges like for an internet connection. Through this, unnecessary users could be removed easily but it requires vast amount of flexibility in the system which could also be upgraded, whenever required.

Another concern is of power consumption. Things like batteries, solar cells are options but they surely can not quench the unending demand for power and thus deemed impractical.

Last but not the least is the issue for proving mobility for such systems by concealing itself in the world. What could be thought for giving a real shape to this model is through the use of something like satellites which will keep track of you without even your knowledge. Consider the example of GPS (Global Positioning System) as an example.

The integration of both complex back-end and front-end is not an easy task and requires 'decentralization' where everyone will be working in a distributed network but with synchronization between each other. Devices like Jini [28] and Universal Plug and Play (UPnP) [29] are paving the way for this third wave of communication. This next wave of communication will have major effect on environment and thus utmost care needs to be given to see that nature is not affected with this creation. Thus, in a nutshell, one must aim at the bigger vision when trying to make a "Halycon Home" for us and through this paper we have tried to achieve the same.

# 7   Conclusion

Every technology is coming into existence to harvest into a boon for the populace, yet it is in the aftermath of its implementation in the world that the darker side of it also comes into picture. One then plans to eliminate it by launching several 'version series' in the market, one after another. In this paper we tried to address drawbacks with the positive side, so as to make a wave of hitting the right target at the first attempt only. Can we do that? Surely, we can by every individual doing his small part to make a 'better world'.  We are in no way against this third era of technology. What we just want is to make a world with no shortcomings, so as not to repent later on. Through this paper we also tried to explain achieving the same by proposing a HALYCON framework and by bringing forth all the points under one roof.

# References

[1] Mark Weiser (1991) *Computer for the 21st Century* http://nano.xerox.com/hypertext/weiser/SciDraft3.html

[2] Mark Weiser and John Seely Brown (1996). *Coming Age of Calm Technology*, Xerox PARC http://www.johnseelybrown.com/calmtech.pdf

[3] *Data Fountain, Money translated to water*, http://www.koert.com/work/datafountain/

[4] Mark Altosaar, Roel Vertegaal, Changuk Sohan, Daniel Chang: Auraorb (2006): *Social Notification Appliance*, Human Media Labs.
*http://interruptions.net/literature/Altosaar-CHI06-p381-altosaar.pdf*

[5] Martin Tomitsch (2007) Towards a Taxonomy for Ambient Information Systems, *Pervasive'07 Workshop* http://deco.inso.tuwien.ac.at/fileadmin/user_upload/Pervasive07-WS-AIS-Taxonomy.pdf

[6] Information Technology Association of America (ITAA), Radio Frequency Identification. RFID ...Coming of age (2004)
 http://www.itaa.org/rfid/docs/rfid.pdf

[7] *Polaris Networks*, http://www.polarisnetworks.net/datasheet/pervcompressrealease230109.pdf

[8] Boaz Carmeli, Benjamin Cohen, Alan J. Wecker (2000) *Proceedings of the eleventh ACM on Hypertext and hypermedia* http://portal.acm.org/citation.cfm?id=336296.336 502

[9] Vannevar Bush (1945) As we may Think, *Atlantic Monthly.*http://www.ps.uni-sb.de/~duchier/pub/vbush/vbush-all.shtml

[10] John Thackara (2001), Pervasive Computing, *Receiver Magazine*, http://www.vodafone.com/flash/receiver/05/articles/pdf/01.pdf

[11] *The Morph Concept*. http://www.nokia.com/about-nokia/research/demos/the-morph-concept

[12] Gene Michael Stover (2005), *Notes about Vannevar Bush's As We May Think* http://cybertiggyr.com/nmemex/nmemex.pdf

[13] Kenneth Arnold: *ConceptNet3*, MIT Media Lab, 2007.  http://conceptnet.media.mit.edu/

[14] F. Hattori, K. Kushima, T. Wasano: *A comparison of Lisp, Prolog, and Ada programming productivity in AI area*, 1985. http://portal.acm.org/citation.cfm?id=319655

[15] John Stasko, Myungcheol Doo, Brian Dorn, Christopher Plaue (2007). *Explorations and Experiences with Ambient Information Systems*, Pervasive'07 Workshop. http://www.cc.gatech.edu/~john.stasko/papers/pervasive07-sys.pdf

[16] John Stasko, Chris Paule, Zach Pausman: *The InfoCanvas- Information Art*, Aware Home, Georgia Tech., http://www.awarehome.gatech.edu/projects/The_InfoCanvas.pdf

[17] Peter Beens: *An Overview of Phidgets – Low Cost USB, Interfacing*, Proceedings of 7th Annual ACSE Conference York University-November, 2005 http://wiki.acse.net/images/b/b6/Phidgets_Peter_Beens_ACSE2005.ppt

[18] *Phidgets: Programming Manual,* http://www.phidgets.com/documentation/Programming_Manual.pdf

[19] Thorsten Prante et al (2003), *HelloWall- Beyond Ambient Displays* , 5[th] International Conference on Ubiquitous Computing, USA.
http://www.ipsi.fraunhofer.de/ambiente/paper/2003/prante-hello.wall_ubicomp03-withCopyright-letter.pdf

[20] Sara Routarinne, Johan Redstrom (2007), *Domestication as Design Intervention* http://www.johan.redstrom.se/papers/domestication.pdf

[21]   *The e-waste problem.* http://www.greenpeace.org/international/campaigns/toxics/electronics/the-e-waste-problem

[22]   *E-waste- An Indian Perspective.* http://www.assocham.org/events/recent/event_64/An_Indian_Perspective_by_Toxic_Links.ppt

[23] *Project Aura,* http://www-2.cs.cmu.edu/~aura/

[24] David A. Cieslikows, Naomi J. Halewood, Kaoru Kimura and Christine Zhen-Wei Quang (2009): *Key Trends in ICT Development, Information and Communication for Development.* http://siteresources.worldbank.org/EXTIC4D/Resources/5870635242066347456/IC4D_2009_Key_Trends_in_ICT_Deelopment.pdf

[25] David Brin: *Transparent Society* http://www.usemod.com/cgi-bin/mb.pl?TransparentSociety

[26] Cheryl Ajluni (2009) *Can Bluetooth and 802.11b co-exist?*   http://www.carl-chapman.com/articles/archives/coexist.pdf

[27] Pervasive Computing: *The Mobile World*, Uwe Hansmann, Springer Professional Computing, pg 395 http://books.google.com/books?id=8yyAbiMPOF0C&printsec=frontcover&dq=pervasive+computing&ei=DBOdSqOcKaCCkASI8PWdAQ#v=onepage&q=&f=false

[28] *Introduction to Jini*, Wikipedia. http://www.jini.org/wiki/Category:Introduction_to_Jini

[29] *Universal Plug and Play*, Wikipedia http://en.wikipedia.org/wiki/Universal_Plug_and_Play